\documentclass[11pt]{article}
\usepackage{epsfig}
\usepackage{graphicx}
\usepackage{url}
\expandafter\let\csname equation*\endcsname\relax
\expandafter\let\csname endequation*\endcsname\relax
\usepackage{amssymb}
\usepackage[latin1]{inputenc}
\usepackage{amsmath}
\usepackage{epsfig}
\usepackage{epstopdf}
\newcounter{fig}
\begin{document}
\title{\ Performance comparison of pulse-pair and wavelets methods for the pulse Doppler weather radar spectrum
}

\author{M. Lagha, M. Tikhemirine, S. Bergheul, T. Rezoug \\
Aeronautical Sciences Laboratory, Aeronautics Departement,  \\ 
Saad Dahlab University, B.P.270, Road of Soumaa, Blida-Algeria
\and and M. Bettayeb \\
Electrical Science Departement, College of Engineering,  \\
University of Sharjah, P.O.Box 27272, Sharjah, United Arabes Emirates
}

\maketitle

\begin{abstract}
In the civilian aviation field, the radar detection of hazardous weather phenomena (winds) is very important. This detection will allow the avoidance of these phenomena and consequently will enhance the safety of flights. In this work, we have used the wavelets method to estimate the mean velocity of winds. The results showed that the application of this method is promising compared with the classical estimators (pulse-pair, Fourier).
\end{abstract}

\section{Introduction}
\label{introduc}

	Mainly in civil aviation, the meteorology plays a big role in the security of flights. Aeronautics depends on reliable information related to the present and future meteorological conditions. This information is given by a meteorological radar. This last detects and locates even distant atmospheric disturbances in form of signals in microwaves.

	Several methods of signal processing are worked out, others offered, to complete successfully the extraction of useful information to be transmitted to the user.

	The pulse-pair method acts on the Doppler radar signal in the temporal domain by autocorrelation. As for the wavelets method, it spreads out in the spectral domain while having references to time.

	The main purpose of this work is to improve the estimates of the average velocity and variances of the detected meteorological disturbances. The use of an algorithm of denoising based on the wavelet transform will lead to this improvement.

   	In section 2, we will introduce the estimation theory of spectral moments. This estimation will be made by two estimators, one temporal, named pulse-pair and another one scalo-temporal, the wavelets. The section 3 introduces results and comments linked to calculations and simulation. Finally a general conclusion will be presented in section 4. 

\section{Estimation theory}
\label{locally}

	The first three (03) moments of the power spectral density of the Doppler spectrum are directly linked up with the desired atmospheric basic parameters. The radar reflectivity (Z), the radial velocity (Vr) and the spectral width of velocities (W). [1], [2], [3], [8].

	Since the return signal in the radar from a range cell (space dimension of an impulse) is generated by the back scattering of a big number of randomly distributed particles and/or by variations of the refraction index of the atmosphere/air, then the process of the received signal is considered (central limit theorem) or approximated by a gaussian random process.

	The shape of the received signal reflected by all particles met by the transmitted beam is: [3], [5], [8].

\begin{equation}
V(t,R)=\sum_{i}^{}A_{i}e^{j2\pi f_{i}\left ( t-2\frac{R_{i}}{c} \right )}.w(t-2\frac{R_{i}}{c})
\end{equation}

with:

	$A_{i} $: Complex amplitude of the individual particle i,

	$f_{i} $: Doppler frequency of the particle i,

	$R_{i} $: Distance from the radar to the particle i,

	$w(t-2\frac{R_{i}}{c}) $: Weighting Function of distance.

	At the $k^{th}$ To the kth radar impulse, the received wave shape may be written in the complex form [3], [5], [6]

\begin{equation}
V_{k}(R)=I_{k}(R)+jQ_{k}(R)
\end{equation}

with:

	$I_{k}(R) $: In phase component.

	$Q_{k}(R) $: Quadrature component.

	The estimation of the meteorological parameters is typically accomplished in the signal processing on the basis of range cells.

\subsection{Pulse-Pair Method}

	The working frequencies of a meteorological radar are in the order of $10^9$ Hz corresponding to wavelength of some centimeters (<10 cm). Generally, the meteorological targets move with speeds lower than 50 m/secs. The Doppler Effect would translate these speeds into a shift of the transmitted frequency. This shift is a few hundred of Hz: much too weak to be directly measured. To reach it, we measure the phase shift between the return of two successive impulses having probed the same volume of space. Indeed, after the return of the second impulse, the target would have changed position which would be translated in a phase shift between both impulses. Hence the pulse-pair method is derived.
\subsection{The pulse-pair technique}

	The classical pulse-pair technique gives the first two spectral moments (speed Doppler and its variance) of the weighted Doppler spectrum of the reflectivity from the function of autocorrelation of the radar complex signal [1], [2], [3], [4], [5].
\begin{equation} \
V[n]=I[n]+jQ[n]
\end{equation}

	Components I and Q are supposed to be independent statistically [5, [6], [7].

	The function of autocorrelation can be expressed for the sequence $V[n]$ of length N and whose samples are spaced by Ts, by: [3], [5], [6]

\begin{equation} \
\hat{R}[1]=\frac{1}{N}\sum_{n=0}^{N-2}V[n+1].V^{*}[n]
\end{equation}

	The mean velocity of a weather target is estimated by [5]:

\begin{equation} \
\hat{v}=-\frac{\lambda}{4\pi T_{S}}arg(\hat{R}[1])
\end{equation}

	And consequently, the spectral width of speeds is done by:

\begin{equation} \
\hat{w}=\frac{\lambda}{2\sqrt{2}\pi T_{S}}\sqrt{ln\left |\frac{R_{S+N}[0]-R_{N}[0]}{R[1]}  \right |}
\end{equation}

	Where index S+N indicates the weather signal merging in the noise N.

\section{Wavelet Method}

	The fundamental assumption in signal processing is to consider the radar echo signal consisting of two parts.

\begin{equation} \
y=f_{a}+\varepsilon 
\end{equation}

	Where $f_{a}$ produced by the Gaussian process of atmospheric diffusion and $\varepsilon$ a noise resulting from various but mainly thermal sources [1].

	In the following, we will deal with the problem by using the wavelets to filter the collected signal then apply the already seen method of processing (pulse-pair) to extract the various spectral moments of interest.

	The primary reasons of the effectiveness of processing by wavelets are linked to the nature of the noise:

		- it is nonstationary

		- it is transitory and of unknown form

	We will suppose that the Gaussian model fits the atmospheric signal $f_{a}$ and noise $\varepsilon$.

\subsection{Discrete wavelet transform}
	The dyadic discrete wavelet transform (DWT) of a signal y(t) is [9], [10]

\begin{equation} \
T_{m,n}=\int_{-\infty }^{+\infty }y(t)2^{-\frac{m}{2}}\Psi ^{*}(2^{-m}t-n)dt
\end{equation}

Where:

	$m$: represents the considered scale.

	$n$: represents the translation of the wavelet $\Psi(t)$.

	$T_{m,n}$: represents a coefficient of correlation between $y(t)$ and $\Psi(t)$.

	$\Psi_{m,n}=2^{-\frac{m}{2}}\Psi ^{*}(2^{-m}t-n)$, represents the orthonormal basis of the mother wavelet$\Psi$.

	Consequently the reconstruction of $y(t)$  from the coefficients $T_{m,n}$ is then:

\begin{equation} \
y(t)=\sum_{m=-\infty }^{+\infty }\sum_{n=-\infty}^{+\infty}T_{m,n}\Psi_{m,n}(t) 
\end{equation}

\subsection{Multiresolution analysis}

	If one samples a continuous signal u with a regular interval unit, one will obtain a sequence with discrete values ${{\rm u}}_{{\rm n}}{\rm =}{\rm u}\left({\rm n}\right){\rm ,\ \ }{\rm n}{\rm \ }\in {\rm \ }{\mathbb Z}$. [11]

	The dwt on one level breaks up the sequence ${{\rm u}}_{{\rm n}}{\rm =}{\rm u}\left({\rm n}\right)$ in two sequences a1 and d1 by a low-pass filter h and a high-pass filter g, both followed by an undersampling of order 2.

	The sequence a1 obtained is called approximation and contains information of low frequencies of ${{\rm u}}_{{\rm n}}{\rm =}{\rm u}\left({\rm n}\right)$ , while the sequence d1 contains information of high frequencies of ${{\rm u}}_{{\rm n}}{\rm =}{\rm u}\left({\rm n}\right)$  called details. h and g are the decomposition filters of finite lengths (FIR).
\begin{equation} \
a^1=\left(u*h\right)\downarrow 2 
\end{equation}
\begin{equation} \
d^1=\left(u*g\right)\downarrow 2 
\end{equation}

	The operation of undersampling by 2 means to take every other sample. It is very useful during the reconstruction operation. This one is carried out by means of filters of reconstruction $\overline{{\rm h}}$ and $\overline{{\rm g}}$. The filters of analysis and those of synthesis are in quadrature mirror (QMF).

	Indeed, the sequence $ u_{n}$  can be found by the expression:

\begin{equation} \
u_n=\left(a^1\uparrow2\right)*\overline{h}+\left(d^1\uparrow2\right)*\overline{g} 
\end{equation}

	The oversampling of order 2 inserts zeros in the sequences in order to find the initial sequence $ u_{n}$ with the same initial number of samples.

	If the operation of decomposition is repeated $J$ times, one obtains the algorithm of filter bank applied to the approximation outputs (see figure 1).

	In the opposite direction, the reconstruction of  $ u_{n}$  is carried out starting from the approximation $ a^{4}$  and the details $d^{1}$, $d^{2}$, $d^{3}$ and $d^{4}$.

\begin{figure}[ht]
\centerline{\includegraphics[width=15.0truecm,clip=]{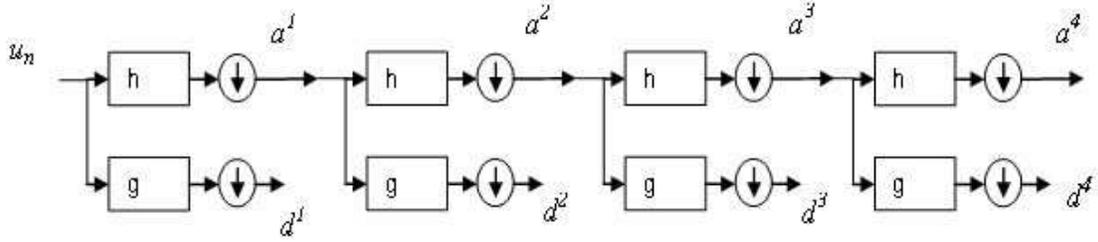}}
\caption{ DWT of $ u_{n}$ to level $ J=4 $: the sequence $ u_{n}$ is decomposed into an approximation $ a^{4}$ and details $d^{1}$, $d^{2}$, $d^{3}$ et $d^{4}$}
\end{figure}

\subsection{Thresholding}
The wavelet coefficients correspond to the details of a signal. A detail can be been ignored without affecting significantly the data.

There exist various methods of thresholding of the wavelet coefficients: hard thresholding and soft thresholding.

Hard thresholding:

\begin{equation} \
d_s=\left\{ \begin{array}{c}
0\ \ \ \ si\ \ \ \left|d\right|<\lambda  \\ 
d\ \ \ \ si\ \ \ \left|d\right|\ge \lambda  \end{array}
\right\} 
\end{equation} \

Soft thresholding or shrinking specific to the complex signals: [12]

\begin{equation} \
d_s=d\left(1-\frac{\lambda }{\left|d\right|}\right) 
\end{equation} \

	As for the various thresholds, one can use:

		- The universal threshold:$\ \ {\lambda }_N=\sqrt{2{\rm log}(Nlog\left(N\right))}$

		- The minimax threshold: (see table I)

		- The SURE threshold: (Stein's unbiased risk estimate).

	Where d,${\rm \ }\lambda $ and N are the wavelet coefficients, the threshold and the length of the signal, respectively.

\noindent \textbf{Table I: Extract of table 1 of [12]}

\begin{tabular}{|p{0.3in}|p{0.6in}|} \hline 
\textbf{N} & \textbf{Minimax (}${{{\mathbf \lambda }}_{{\mathbf N}}}^{{\mathbf *}}$\textbf{)} \\ \hline 
16 & 1.763 \\ \hline 
128 & 1.973 \\ \hline 
256 & 2.176 \\ \hline 
512 & 2.371 \\ \hline 
1024 & 2.560 \\ \hline 
2048 & 2.741 \\ \hline 
\end{tabular}

\subsection{Signal processing algorithm}
	The thresholding is efficient means to ignore the weakest details, compared to the selected threshold ,${\rm \ }\lambda $, which one can compare to the noise and preserve only the most important wavelet coefficients. One rebuilds, then, the signal from the remaining coefficients which represent the denoised data (see figure 2).

\begin{figure}[ht]
\vspace*{.5cm}
\centerline{\includegraphics[width=10.0truecm,clip=]{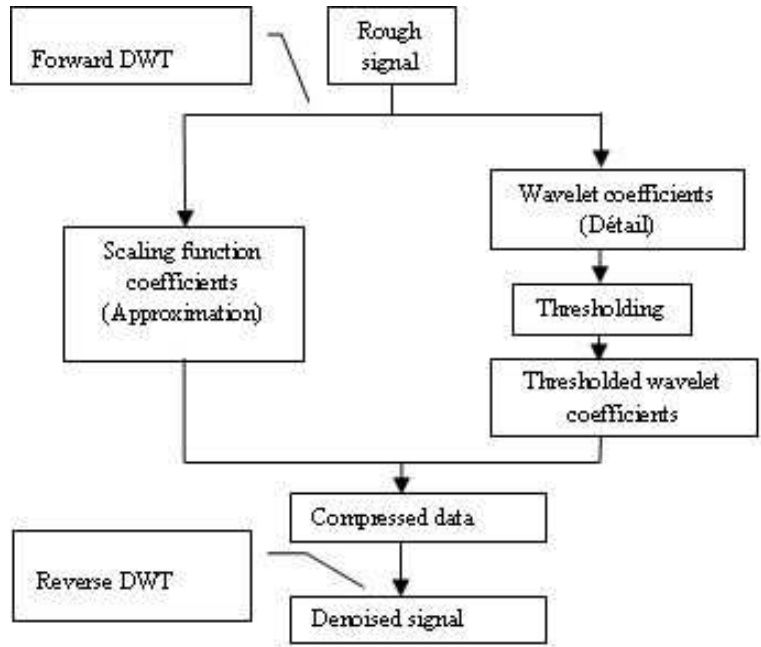}}
\caption{ Synoptic diagram of the wavelet denoising method.}
\end{figure}

\section{Results and comments}

	In this work, we have generated data representing the echoes of a weather radar of wind disturbance.

	For that, we have used the algorithm of Zrniç developed in 1975 and included in works of R.D. Palmer. These data consist of signals I and Q. These calculations were carried out on a set of ten range cells taken on one radar azimuth.
\begin{figure}[ht]
\centerline{\includegraphics[width=8.0truecm,clip=]{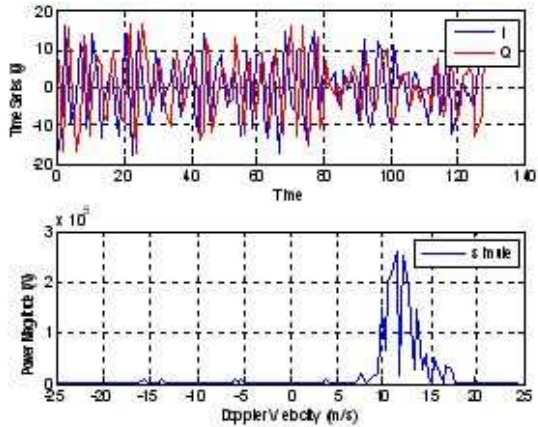}}
\caption{ signals I and Q and Doppler spectrum of the cell $N1$}
\end{figure}

	In figure 3, a, b, we have represented signals I and Q and their spectra respectively for the range cells 1 and 5. As one can see it on the figure, this spectrum is a Gaussian form.

	The estimates of the mean velocity and the spectral width of the weather disturbances already generated above are represented on figures 4 and 6.
\begin{figure}[ht]
\centerline{\includegraphics[width=8.0truecm,clip=]{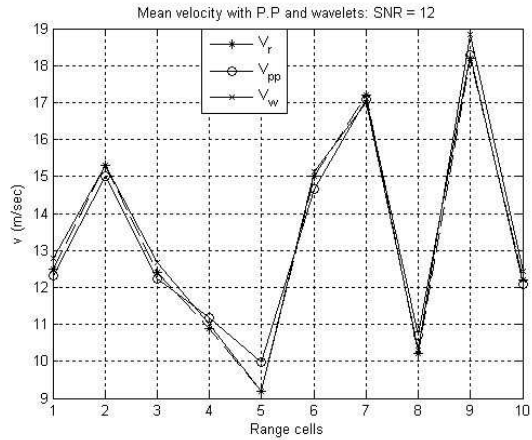}}
\caption{ Estimate of mean velocities of the wind by the Pulse-pair and wavelet estimators.}
\end{figure}

	It is noticed that the estimate of the mean velocity of the wind by the pulse-pair algorithm is close to the real speed. On the other hand, the estimate by the method of Fourier is less close.

	We also notice that the estimation of the spectral width is less important with the wavelets than with the pulse-pair algorithm.
\begin{figure}[ht]
\centerline{\includegraphics[width=8.0truecm,clip=]{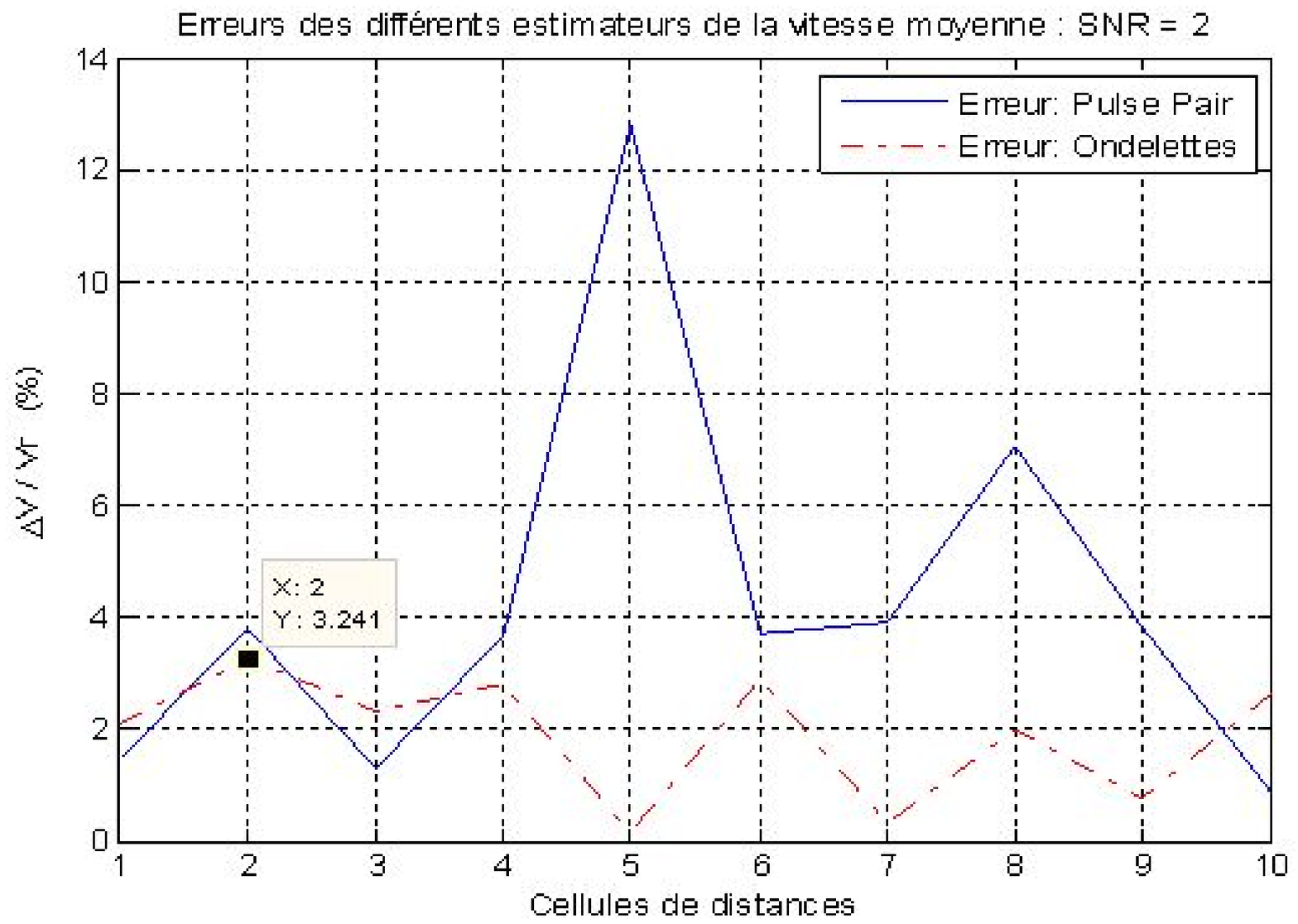}}
\caption{ Errors on the estimate of mean velocities of the wind by the Pulse-pair and wavelet estimators.}
\end{figure}

\begin{figure}[ht]
\centerline{\includegraphics[width=8.0truecm,clip=]{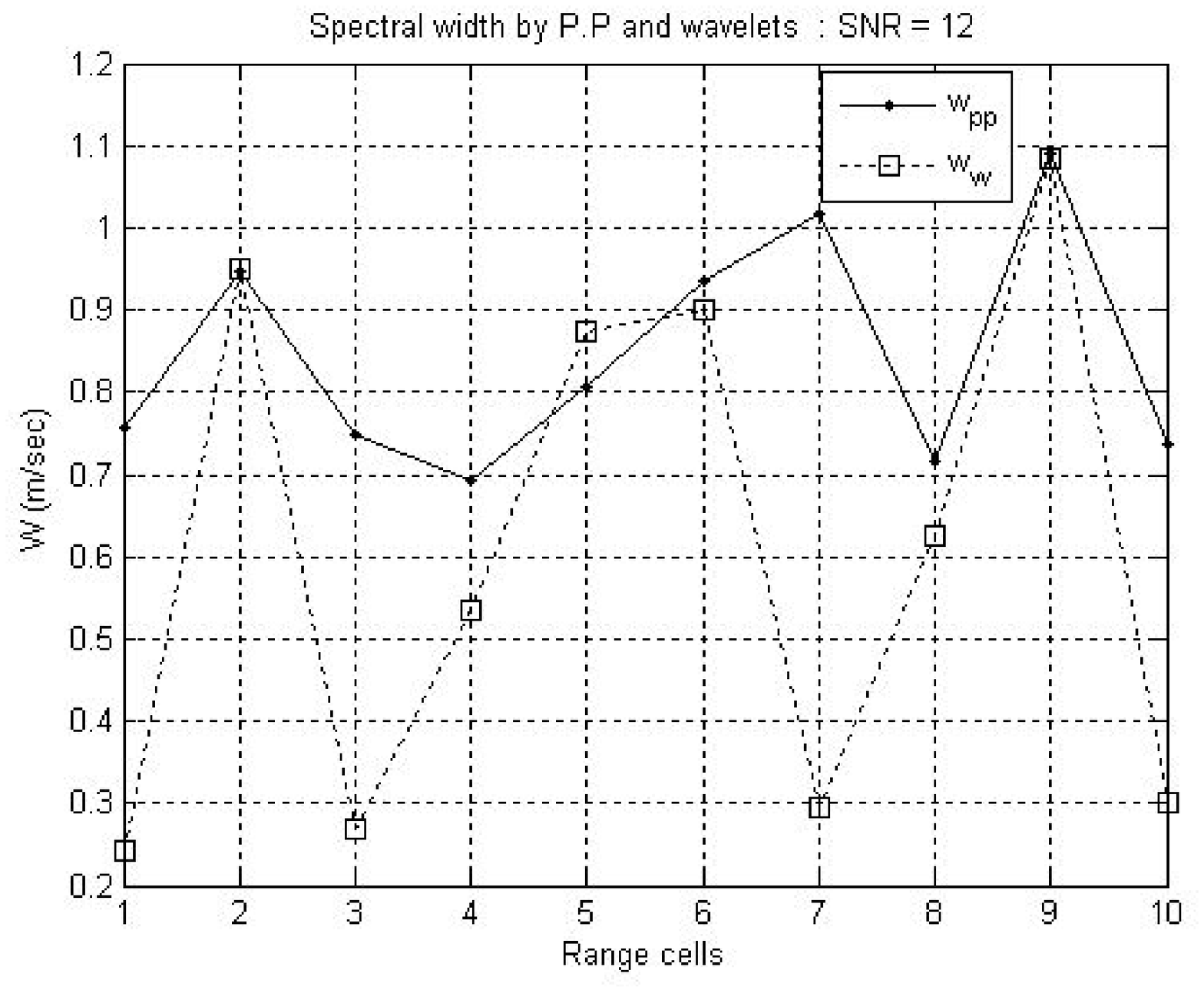}}
\caption{ Spectral width obtained the Pulse-pair and wavelet estimators.}
\end{figure}

	In addition, by applying the wavelet method, we have obtained better results compared to Fourier and pulse-pair, see figure 4. This indicates the interest to have used this method. These insufficiencies of the algorithm of Fourier are inherent in his approach and they are due to:

		- Resetting the frequencies which are apart from the field of work, which generates spectral losses.

		- Spectral resolution (close frequencies).

\section{Conclusion}
\label{concl}

	In this work, we have re-examined the estimate of the spectral moments of the signals received by a Doppler radar, namely the moments of orders one (mean velocity of the wind) and two (spectral width).

	It comes out from it, through the results obtained, that the algorithm of denoising by the means of the wavelet transform of the signal to be treated is efficient. Our results were consolidated by the comparison of the errors of the various used algorithms: pulse-pair and wavelet. The proof is the reduction of the relative error made on the parameters estimated compared to the input data compared to the results obtained by the pulse-pair algorithm as well as the reduction of the spectral width

\vskip .5cm 

{\bf Acknowledgment} 
	This work was realized under the collaboration between the Aeronautical Science Laboratory of Blida University and the Search group of the College of Engineering, Sharjah, UAE.


\vskip .1cm


\begin{thebibliography}{10}



\bibitem {Pazmany} A. L. Pazmany, J. C. Galloway, {\em Polarization Diversity Pulse-Pair Technique for Millimeter-Wave Doppler Radar Measurements of Severe Storm Features}, Journal of Atmospheric and Oceanic Technology, jan. 1999.
\bibitem{Aalfs} D. D. Aalfs, E. G. Baxa, E. M. Bracalente, {\em Signal Processing Aspects of Windshear Detection}, Technical Feature, Microwave Journal, Sep. 1993, pp.76-96.
\bibitem{Doviak} R.J. Doviak, D. S. Zrnic,{\em Doppler weather radar}, Proceedings of the IEEE, Vol. 67, No. 11, November 1979.
\bibitem{Serafin} R. J. Serafin, {\em Meteorological radar}, Chapter 23, Radar Handbook, McGraw-Hill Book Company, 2nd edition1990, pp.1-33.
\bibitem{Brigni} V. N. Bringi, V. Chandrasekar, {\em Polarimetric Doppler weather radar}, Cambridge University Press 2004.
\bibitem{Lagha} M. Lagha, M. Bensebti, {\em Performances comparison of pulse-pair and 2-step prediction algorithms for the Doppler spectrum}, Multidimensional Systems and Signal Processing, 2008.
\bibitem{Lagha} M Lagha, M Bensebti, {\em Performance Comparison of Pulse-Pair and 2-step Prediction Approach to the Doppler Estimation}, - International Symposium on Industrial Electronics, 2006.
 \bibitem {Lagha} M Lagha, M Bensebti, {\em  Doppler Spectrum Estimation By Ramanujan-Fourier Transform (RFT)}, - Digital Signal Processing, 2009 - Elsevier.
 \bibitem {Mallat} St\'ephane Mallat, {\em A Wavelet Tour Of Signal Processing}, Academic Press is an imprint of Elsevier, 3rd Edition 2009.
 \bibitem {Daubechies} I. Daubechies, {\em Ten lectures on wavelets},  SIAM, 1992.
 \bibitem {Justen} L. A. Justen, G. Teschke, V. Lehmann.,{\em  wavelet-based methods for clutter removal from radar wind profiler data}. Proceedings of SPIE, 2003.
 \bibitem {Sardy} S. Sardy. {\em minimax threshold for denoising complex signals with waveshrink}. IEEE Transactions on Signal Processing, 2000.


\end{thebibliography}
\end{document}